\documentclass[12pt]{article}
\usepackage{epsfig}
\usepackage{cite}
\begin{document}

\begin{titlepage}

\hfill IFIC/11-33

\hfill FTUV-11-301

\vspace{1.5cm}

\begin{center}
\ \\
{\bf\large Monopolium detection at the LHC}
\\
\date{ }
\vskip 0.70cm

Luis N. Epele$^{a}$, Huner Fanchiotti$^{a}$, Carlos A. Garc\'{\i}a
Canal$^{a}$, Vasiliki A. Mitsou$^{b}$\\ and \\Vicente Vento$^{b,c}$

\vskip 0.30cm

{(a) \it Laboratorio de F\'{\i}sica Te\'{o}rica, Departamento de
F\'{\i}sica, IFLP \\ Facultad de Ciencias Exactas, Universidad
Nacional de La Plata\\
C.C. 67, 1900 La Plata, Argentina.}\\
({\small
E-mail: epele@fisica.unlp.edu.ar, cgarciacanal@fisica.unlp.edu.ar,
huner@fisica.unlp.edu.ar})

\vskip 0.3cm

{(b) \it Instituto de
F\'{\i}sica
Corpuscular (IFIC)\\
Universidad de Valencia and CSIC\\
Parc Cient\'{\i}fic, Apartado de Correos 22085, E-46071 Valencia, Spain.}\\
({\small E-mail:
vasiliki.mitsou@ific.uv.es})\vskip 0.3cm

{(c) \it Departamento de F\'{\i}sica Te\'orica \\
 Universidad de Valencia\\
 E-46100 Burjassot (Valencia), Spain.} \\ ({\small E-mail:
vicente.vento@uv.es}) \vskip 0.3cm

\end{center}

\vskip 1cm \centerline{\bf Abstract}

Dirac monopoles have been widely studied and searched, though never found. A way out of this impasse is
the idea that monopoles are not seen freely because they are confined by their
strong magnetic forces forming a  monopole-antimonopole bound state called monopolium.
Monopolium  was shown to be produced abundantly and in some scenarios easier 
to detect than monopoles themselves. The Large Hadron Collider is reaching energies never achieved 
before allowing the  search for exotic particles in the TeV mass range. We extend 
our previous analysis to the observability of monopolium at LHC in the $\gamma \, \gamma$ channel
 particularizing our quantitative discussion to monopolium masses that can be detected with integrated luminosites $\sim 1$ fb$^{-1}$.

 \vspace{1cm}

\noindent Pacs: 14.80.Hv, 95.30.Cq, 98.70.-f, 98.80.-k


\end{titlepage}

\section{Introduction}

Numerous experimental searches for magnetic monopoles have been carried out
but all have met with failure\cite{Craigie:1986ws,Martin:1989ms,Abbott:1998mw,Eidelman:2004wy,Mulhearn:2004kw,
Giacomelli:2005xz,Abulencia:2005hb,Milton:2006cp,Yao:2006px, 
Balestra:2011ks}. These experiments have led to a lower mass limit in the range from 
$350-500$ GeV.

This lack of experimental confirmation has led many physicists to
abandon the hope in their existence. A way out of this impasse is
the old idea of Dirac \cite{Dirac:1931kp,Dirac:1948um,Zeldovich:1978wj}, namely,
monopoles are not seen freely because they are confined by their
strong magnetic forces forming a bound state called monopolium
\cite{Hill:1982iq,Dubrovich:2002gp}. This idea was the leitmotiv
behind our research, namely we proposed that monopolium,  due to its 
 bound state structure,  might be easier to detect than free monopoles  
\cite{Epele:2007ic,Epele:2008un}. We showed that certain 
parameterizations of the mass and the width,
allowed for such scenario.

The Large Hadron Collider (LHC), which entered last year  in operation colliding  $3.5$ TeV protons,
will probe the energy frontier opening possibilities for new physics including the discovery of magnetic monopoles either
directly, a possibility contemplated long time ago
\cite{Ginzburg:1981vm,Ginzburg:1982fk}, and revisited frequently \cite{Abbott:1998mw,Abulencia:2005hb,
Kurochkin:2006jr,Epele:2011cn,Dougall:2007tt,Kalbfleisch:2000iz,Ginzburg:1998vb,Ginzburg:1999ej}, 
or through the discovery of monopolium, as advocated in refs. \cite{Epele:2007ic,Epele:2008un}. 
The direct observation is based on the search for highly ionizing massive particles at the ATLAS~\cite{atlas}
and CMS~\cite{cms} detectors or at the MoEDAL experiment~\cite{moedal}, the design of which is
optimized to search precisely for such exotic states.
These developments motivate our present research which analyzes the production of monopolium at
LHC by the mechanism of photon fusion and its subsequent decay into $\gamma\, \gamma$.

As was shown in our previous work, photon fusion is ideal for monopolium production \cite{Epele:2008un} . 
To complete the analysis, we study here how monopolium,   produced at the collision point, can be detected through its
decay in a pair of extremely energetic photons, a channel for which LHC detectors are optimized,  since  this is one of the decay 
channels for finding a light Higgs with low counting rates.

In the next section we establish the formalism for monopolium production and decay.
Section 3 describes  how to incorporate the elementary process into $ p-p$ scattering. 
Section 4 presents our results and studies the variation with the two parameters, monopole mass and monopolium mass.
Finally in section 5 we draw some general conclusions of our study.

\section{Monopolium dynamics}

For the case of monopole interactions at energies higher than their mass  there is no universally accepted  
effective field theory\cite{Schwinger:1966nj,Zwanziger:1970hk,Gamberg:1999hq}.  However, the study of  the  classical interaction of a monopole passing by a 
fixed electron  leads to an interaction for the monopole which is associated with the electric field 
felt. If one compares this interaction with that of  a positron passing by an electron, one realizes 
that the difference between QED and dual QED \cite{Gamberg:1999hq}, the theory without strings for  monopole interactions, 
is simply to change the electric charge by the magnetic 
charge times the velocity. For this reason we will
employ a minimal model of monopole interaction which assumes an effective
monopole photon-coupling which is proportional to the monopole's
induced electric field $g\beta$ for a monopole moving with velocity
$\beta$ \cite{Mulhearn:2004kw,Epele:2011cn,Dougall:2007tt,Kalbfleisch:2000iz}.

The Dirac quantization condition does not specify the spin of the
monopoles. We choose here monopoles of spin $1/2$  
coupled in monopolium
of spin $0$ in order to have a minimum energy radial structure.


\begin{figure}[htb]
\centerline{\epsfig{file=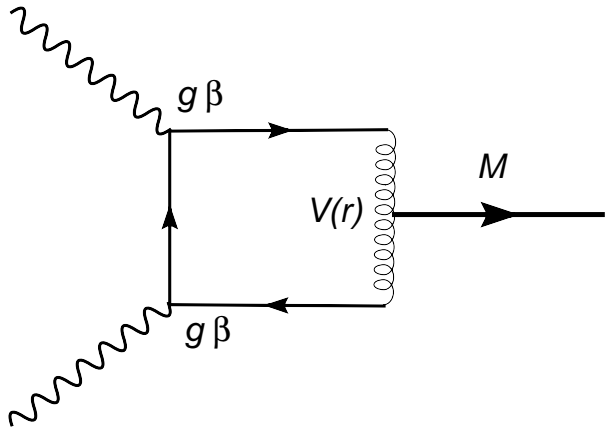,width=8cm,angle=0}}
\caption{\small{Diagrammatic description of the elementary
subprocess of the monopolium production from photon fusion. 
$V(r)$ represents the interaction binding the monopole-antimonopole pair to form monopolium.}}
\label{gbetab}\end{figure}


Recently we studied the production of monopolium by photon 
fusion at LHC \cite{Epele:2008un}. 
The elementary subprocess calculated is shown in Fig.
\ref{gbetab}. The
standard expression for the cross section of the elementary
subprocess for producing a monopolium of mass $M$ is given by


\begin{figure}[htb]
\centerline{\epsfig{file=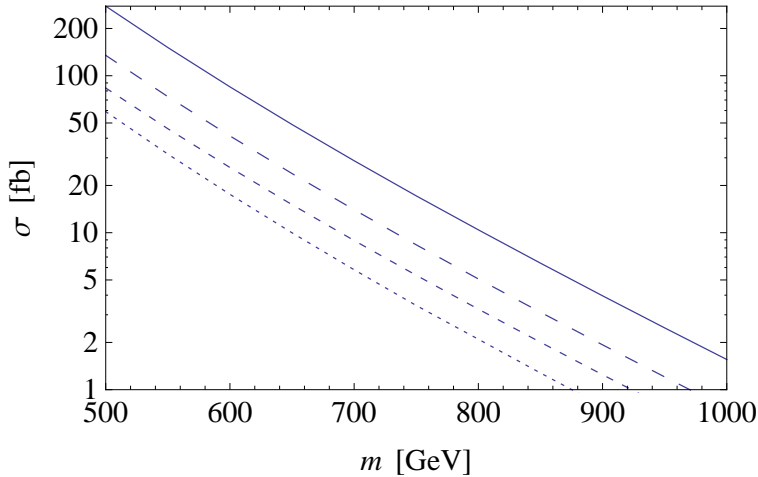,width=10cm,angle=0}}
\caption{\small{Total cross section for monopolium production at LHC with $3.5$ TeV beams for monopole masses ranging from $500$ to $1000$ GeV (full curve). The broken curves represent the different contributions to the total cross section as described in section 3: semielastic (dashed), elastic (shorst dashed) and inelastic (dotted). We have chosen 
a binding energy $\sim 2\;m/15$ and  $\Gamma_M = 10$ GeV. }}
\label{Mproduction}\end{figure}

\begin{equation}
\sigma (2 \gamma \rightarrow M) = \frac{4\pi}{E^2}  \frac{M ^2
\,\Gamma (E) \, \Gamma_M}{\left(E^2 - M^2\right)^2 +
M^2\,\Gamma_M^2}, \label{ppM1}
\end{equation}
where  $\Gamma (E)$, with $E$ off mass shell, describes the production
cross section. Note that $\Gamma[M] = 0$. $\Gamma_M$ arises from the softening of the
delta function, $\delta(E^2 - M^2)$ and therefore is, in principle,
independent of the production rate $\Gamma (E)$ and can be attributed to the beam 
width \cite{Jauch:1975sp,Peskin:1995hc}.

In Fig. \ref{Mproduction} we show the total cross section for monopolium production from photon fusion
under present LHC running conditions for a monopole mass ($m$)  ranging from $500$ to $1000$ GeV. 
In the figure the binding energy is fixed for each mass  ($2\;m/15$), chosen so that for our case study, $m=750$ GeV, the binding energy is
$100$ GeV and thus $M=1400$ GeV. With this choice the monopolium mass ($M$) ranges from $933$ to $1866$ GeV. 
We notice that detection would be possible with an integrated luminosity of $1$ fb$^{-1}$ if the chosen binding energy
is at the level of $10\% $ of the monopole mass or higher. In the present analisis we study  binding energies small compared to the bound state mass, $M$, in order to be consistent  with the formalism 
used.

\begin{figure}[htb]
\centerline{\epsfig{file=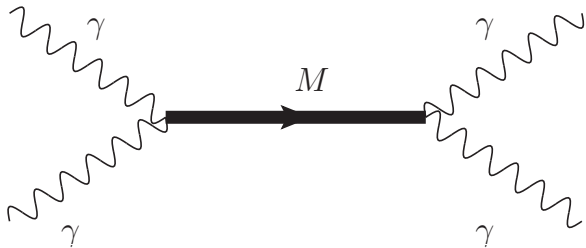,width=8cm,angle=0}}
\caption{\small{Diagrammatic description of the monopolium production and decay.}}
\label{ggMgg}\end{figure}

The interest in this paper is in the detection of photons after monopolium decay. 
The elementary subprocess is shown in Fig.
\ref{ggMgg}, which could be considered as a contribution to light-by-light scattering
in the presence of  monopolium.

The standard expression for the cross section of this elementary
subprocess, after having integrated over angles, is given by 

\begin{equation}
\sigma (\gamma\, \gamma \rightarrow M \rightarrow \gamma \, \gamma ) = \frac{4\pi}{E^2}  \frac{M ^2
\,\Gamma^2 (E)}{\left(E^2 - M^2\right)^2 +
M^2\,\Gamma_M^2}. \label{ppM2}
\end{equation}
Here  $\Gamma (E)$, 
with $E$ off mass shell, describes the vertex $\gamma\, \gamma \, M$.
Monopolium is stable in the center of mass but we add an experimental Gaussian  width $\Gamma_M \sim 10$ 
GeV  in line with the values used in ref. \cite{Allanach:2000nr}. 
\vskip 0.5cm


\begin{figure}[htb]
\centerline{\epsfig{file=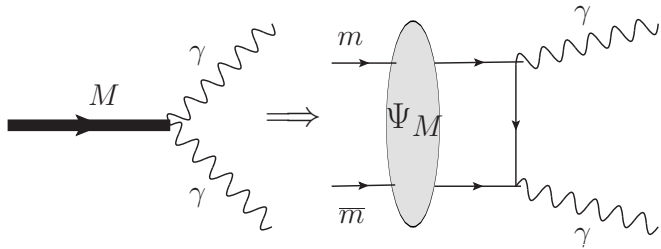,width=9cm,angle=0}}
\caption{\small{The monopolium vertex and its microscopic description
.}}
\label{Mggvertex}\end{figure}

\vskip 0.5cm
We recall now the computation of the $\Gamma (E)$, which represents
the vertex of the monopolium decay to $\gamma\, \gamma$ . The calculation,
following standard field-theory techniques of the decay of a
non-relativistic bound state,
leads to

\begin{equation}
\Gamma (E) =
\frac{32\,\pi\,\alpha_g^2}{M^2}\,\left|\psi_M(0)\right|^2 .
\end{equation}
We have used the conventional approximations for this calculation:  the monopole and antimonopole,  
forming the bound state, are treated as on-shell particles,  when calculating the elementary scattering process shown
on the right of Fig. \ref{Mggvertex}; the bound state is described by a wave function obtained from a Coulomb-type 
interaction between the pair \cite{Epele:2008un,Jauch:1975sp,Peskin:1995hc}. However, 
once the calculation is performed  we substitute $2m$ by $M$, where $m$ is the monopole
 mass 
 to take into account  binding. In the expression, 
 $\alpha_g$ corresponds to the photon--monopole
coupling and $\psi_M$ is the monopolium ground state wave function.
\vskip 0.5cm

\vskip 0.5cm

Before we proceed let us raise a warning on our calculation. Bound 
relativistic systems are notoriously difficult dynamical objects. We proceed here
by performing a non-relativistic calculation of the monopolium wave function. 
The validity of the non-relativistic approximation in the bound state wave functions 
was analyzed in a previous work \cite{Epele:2007ic}. In the dynamics of the decay 
formula the substitution of $2m$ by $M$ constitutes an intuitive way 
to take the off-shellness of the monopoles, i.e. their binding energy in monopolium,  into account. 
For the purposes of estimation both approximations seem reasonable specially
since our binding energies will never exceed $15\%$ of the monopole mass, i.e. less than $10\%$ 
of the total mass of the system.

Using the Coulomb wave functions of ref.\cite{Epele:2007ic}
expressed in the most convenient way to avoid details of the
interaction, which will be parameterized by the binding energy, one
has

\begin{equation}
|\psi_{M}(0)|^2 = \frac{1}{\pi}\left(2 - \frac{M}{m}\right)^{3/2}\; m^3,
\end{equation}
and the effective monopole coupling theory described above 
in the case of monopolium
production, gives rise to \cite{Epele:2008un},

\begin{equation}
\frac{\Gamma(E)}{M}= 2\left(\frac{ \beta^2}{\alpha}\right)^2 \left(\frac{m}{M}\right)^3  \left(2- \frac{M}{m}\right)^{3/2}
m^3.
\end{equation}
Here, $\alpha$ is the fine structure constant and $\beta$ the
monopolium velocity,

\begin{equation}
\beta= \sqrt{1-\frac{M^2}{E^2}} ,
\end{equation}
which is the velocity of the monopoles moving in the monopolium
system.

Note that due to the value of $\beta$ the vertex vanishes at the
monopolium mass, where the velocity is zero. Therefore a static
monopolium is stable under this interaction. We refer to refs. 
\cite{Mulhearn:2004kw,Kalbfleisch:2000iz} for a thorough discussion 
on Lorentz invariance of the theory.

A caveat is due here. There is a duality of treatments in the above
formulation as can be seen in Fig. \ref{gbetab}. The static coupling is treated
as a Coloumb like interaction of coupling $g$ binding the monopoles
into monopolium, although ultimately the details are eliminated in
favor of the binding energy parameterized by the monopolium mass
$M$. We find in this way a simple parametric description of the
bound state. The dynamics of the production of the virtual
monopoles, to be bound in monopolium, is described in accordance
with the effective theory \cite{Kalbfleisch:2000iz,Epele:2011cn}, and
this coupling is $\beta g$. This is similar to what is done in heavy
quark physics \cite{Pennington:2005ww}(see his figure 5), where the
wave function is obtained by a parametric description using
approximate strong dynamics while the coupling to photons is
elementary.


\begin{figure}[t]
\epsfig{file=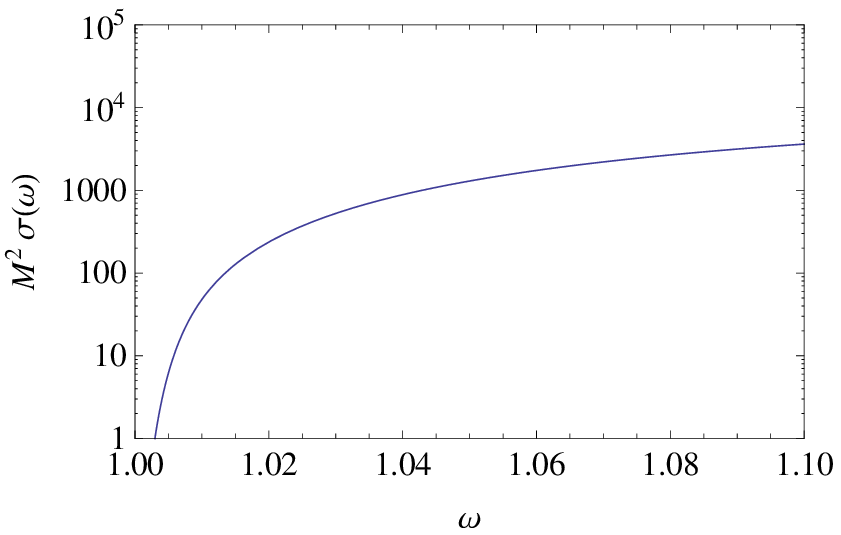,width=6.5cm,angle=0} \hspace{1cm}
\epsfig{file=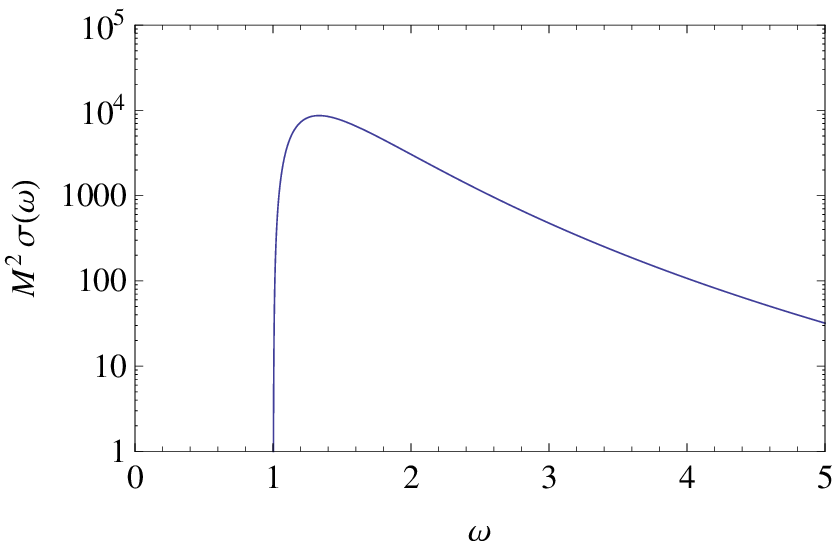,width=6.5cm,angle=0}
\caption{\small{Elementary cross section as a function of $\omega = E/M$ in units of $1/M^2$ calculated for $m=750$ GeV and $M= 1400$ GeV. Left: near threshold. Right:  away from threshold.}} \label{xsec}\end{figure}


The production cross section can now be written as,

\begin{equation}
M^2\;\sigma(\gamma\,\gamma \rightarrow M\rightarrow\gamma\,\gamma ) = 16 \;\pi
 \left(\frac{\beta}{\alpha}\right)^4  \left(\frac{m}{E}\right)^6\frac{(2 - \frac{M}{m})^3}{1 + \frac{M^2 \Gamma_M^2}{E^4 \beta^4}}.
\label{ggxsecM}\end{equation}

The above cross 
section satisfies comfortably the unitarity limit 
\cite{Milton:2008pn},
\begin{equation}
\sigma \le \frac{\pi}{3 E^2}.
\end{equation}

To feel safe with our approximations we consider the binding energy much smaller than $m$, i.e. $M \sim 2m$. In this case the elementary cross section has two very different behaviors as shown in  Fig. \ref{xsec} : i) at threshold it is dominated by $\beta$ and the cross section rises (see left figure); ii) away from theshold the dominant behavior is the $1/E$ dependence and the cross section drops faster than the unitary limit. The conflict between these two behaviors produces a  wide bump-like structure.

\section{Analysis of p-p scattering}

\begin{figure}[t]
\begin{center}   
\includegraphics[scale=0.5]{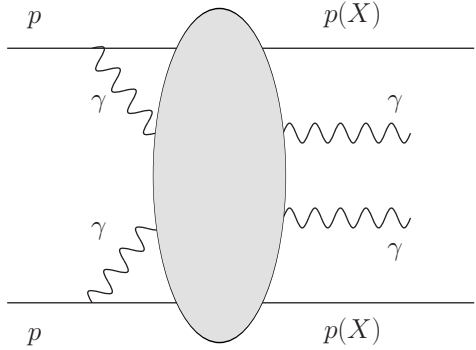}
\caption{Processes contributing to the $\gamma \gamma$ cross section. The blob contains the three cases described in the text. $p$ represents the proton and $X$ any other hadronic state.}
\label{pp} 
\end{center}
\end{figure}

LHC is a proton-proton collider, therefore, in order to describe the production and desintegration of the monopole-antimonopole pair, 
we have to study the following processes above the monopole threshold ($ \beta > 0 \rightarrow E \geq M $),

\begin{eqnarray}
p + p &  \rightarrow &  p(X) + p(X) + \gamma + \gamma,
\end{eqnarray}
shown globally in Fig.\ref{pp}, where $p$ represents the proton, $X$ an
unknown final state and we will assume that the blob is due exclusively to monopolium. This diagram summarizes
the three possible processes:

\begin{itemize}

\item [i)] inelastic $p+ p \rightarrow X+X + \gamma +\gamma
\rightarrow X + X +M \rightarrow  X + X + M +\gamma + \gamma$

\item [ii)] semi-elastic $ p + p \rightarrow p + X + \gamma + \gamma
\rightarrow p + X + M \rightarrow  p + X + M +\gamma + \gamma$

\item [iii)] elastic $p + p \rightarrow p + p + \gamma + \gamma
\rightarrow p + p +  M \rightarrow  p + p  + M + \gamma + \gamma$.
\end{itemize}

In the inelastic scattering, both intermediate photons are radiated
off partons (quarks or  antiquarks) in the colliding protons.

In the semi-elastic scattering one intermediate photon is radiated
by a quark (or antiquark), as in the inelastic process, while the
second photon is radiated from the other proton, coupling to the
total proton charge and leaving a final state proton intact.

In the elastic scattering both intermediate photons are radiated
from the interacting protons leaving both protons intact in the
final state.

In the blob we incorporate the elementary subprocess shown in Fig.
\ref{ggMgg} and  described by Eq. (\ref{ggxsecM}).

We calculate the $\gamma \gamma$ fusion for monopolium production
following the formalism of Drees et al. \cite{Drees:1988pp,Drees:1994zx}.

In the inelastic scattering, $p + p\rightarrow X+ X + \gamma +
\gamma \rightarrow  X +X + M+ \gamma + \gamma$, to approximate the quark distribution
within the proton we use the Cteq6--1L parton distribution functions
\cite{CTEQ} and choose $Q^2 = \hat{s}/4$ throughout, where $\hat{s}$ is the center of
mass energy of the elementary process.

We employ an equivalent--photon approximation for the photon
spectrum of the intermediate quarks \cite{Williams:1934ad,vonWeizsacker:1934sx}.

In semi--elastic scattering, $p + p\rightarrow p+ X+ \gamma +  \gamma
\rightarrow p+ X + M + \gamma + \gamma $, the photon spectrum associated with the
interacting proton must be altered from the equivalent--photon
approximation for quarks to account for the proton structure.  To
accommodate the proton structure we use the modified
equivalent--photon approximation of \cite{Drees:1994zx}.

\begin{figure}[htb]
\begin{center}   
\includegraphics[scale=1.2]{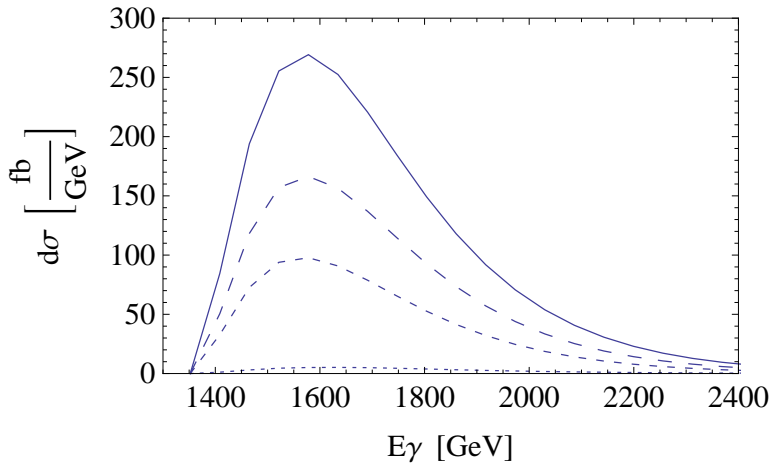}
\caption{The differential cross section (solid line) and its components (semielastic (dashed), elastic (short dashed) and inelastic (dotted)) 
as a function of the center-of-mass photon energy for a monopole mass of $750$ GeV and a monopolium mass of $1400$ GeV. }
\label{ppMggXsec} 
\end{center}
\end{figure}

The total cross section is obtained 
as a sum of the three processes. The explicit expressions for the
different contributions can be found  in \cite{Dougall:2007tt}.

In order to obtain the differential photon-photon cross section from the above
formalism we develop a procedure  which we exemplify with the elastic 
scattering case. In that case the $p-p$ cross section is given by \cite{Drees:1994zx},

\begin{equation}
\sigma_{pp} (s)  = \int_{M^2/s}^1 d z_1  \int_{M^2/s z_1}^1 d z_2 f(z_1) f(z_2) \sigma_{\gamma \gamma} (z_1 z_2  s),
\end{equation}
where $\sqrt{s}$ is the center of mass energy of the $pp$ system and the $f$'s represent the elastic photon spectrum.

We perform the following change of variables 
$$ v= z_1 z_2  \;  , \; w= z_2 \; ,$$
which leads to 
\begin{equation}
\sigma_{pp} (s)  = \int_{M^2/s}^1 d v  \int_{v}^1 \frac{d w}{w}  f(\frac{v}{w}) f(w) \sigma_{\gamma \gamma} (v s). 
\end{equation}
Note that to fix the center of mass energy of the photons is equivalent to fix $v$. For fixed $v$ we have,

\begin{equation}
\frac{d \sigma_{pp}}{dv} (s)  = \int_{v}^1  \frac{d w}{w}  f(\frac{v}{ w}) f(w) \sigma_{\gamma \gamma} (v s),
\end{equation}
which can be rewritten in terms  of $E_\gamma$, the center of mass energy  of the photons, and the elementary photon-photon cross section as,

\begin{equation}
 \frac{d \sigma_{pp}}{dE} (E_\gamma) = \frac{2 E_\gamma}{s}  \; \sigma_{\gamma \gamma} (s_{\gamma \gamma})\;\int_{s_{\gamma  \gamma}/s}^1  \frac{d w}{w}  f(\frac{s_{\gamma  \gamma}}{ w}) f(w).  
\end{equation}
This procedure can be generalized easily to the semielastic and inelastic cases, where the appropriate change of variables are

$$ v= z_1 z_2 x_1  \; , \;  w= z_2 x_1 \; , \;  u =  x_1$$
and
$$ v= z_1 z_2 x_1  x_2 \; , \;  w= z_2 x_1 x_2 \; , \;  u =  x_1 x_2 \; , \; t = x_2,$$
respectively which one has to introduce into integral expressions with  a product of three $f$'s (semielastic) or four $f$'s (inelastic) representing  quark densities and photon spectrum \cite{Drees:1988pp,Drees:1994zx}.

\section{Results and Discussion}

Our aim is to show scenarios which could arise during the present  LHC running period and to discuss general properties 
of the monopolium system which might serve when higher luminosities are achieved.  In Fig. \ref{ppMggXsec} we show the 
structure for the differential cross section. It is a wide bump, starting very close after threshold, i.e. the 
monopolium mass ($1400$ GeV in this case), and extending for about $1000$ GeV . We show in the figure the contribution of the different 
components to the cross section.  The elastic and semielastic components dominate. The behavior is well understood by the structure of Eq. \ref{ggxsecM}, 
the bump initiates due to the rising of the cross section close to threshold associated with its $\beta$ behavior. Close to threshold  $\beta$ takes
almost its asymptotic value of 1 and the $1/E$ behavior of the cross section starts to softly dominate  (recall Fig \ref{xsec}).

\begin{figure}[htb]
\begin{center}   
\includegraphics[scale=1.2]{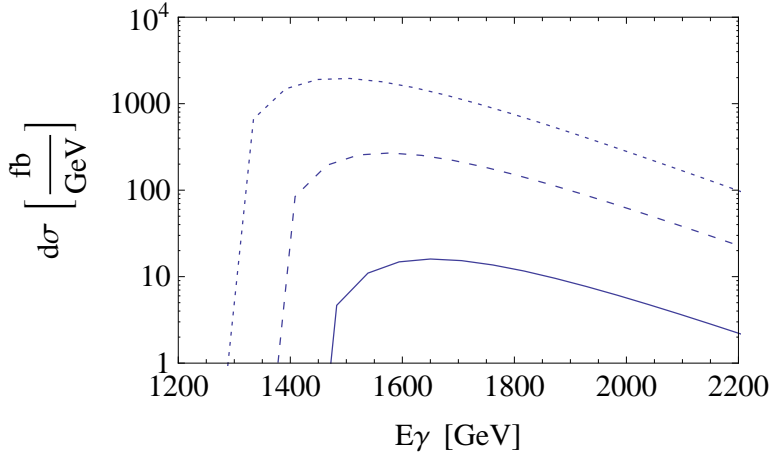}
\caption{Differential two photon cross section as a function of the photon center-of-mass energy for a monopole mass of $750$ GeV 
and different binding energies for monopolium: solid line $75$ GeV, dashed $150$ GeV and solid $225$ GeV. }
\label{MggEbind} 
\end{center}
\end{figure}

\begin{figure}[htb]
\begin{center}   
\includegraphics[scale=1.2]{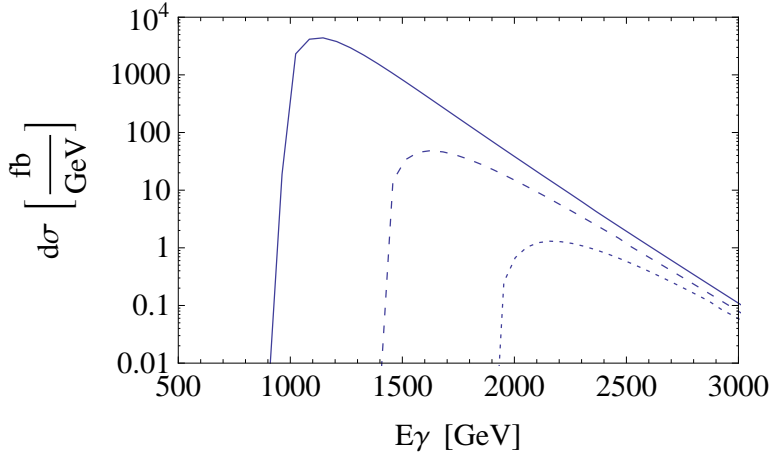}
\caption{Differential cross section as a function of photon energy for different monopole masses: $500$ GeV (solid), $750$ GeV (dashed), $1000$ GeV (dotted) and fixed binding energy ($100$ GeV).}
\label{Mmonopolemasses} 
\end{center}
\end{figure}

\begin{figure}[htb]
\begin{center}   
\includegraphics[scale=1.4]{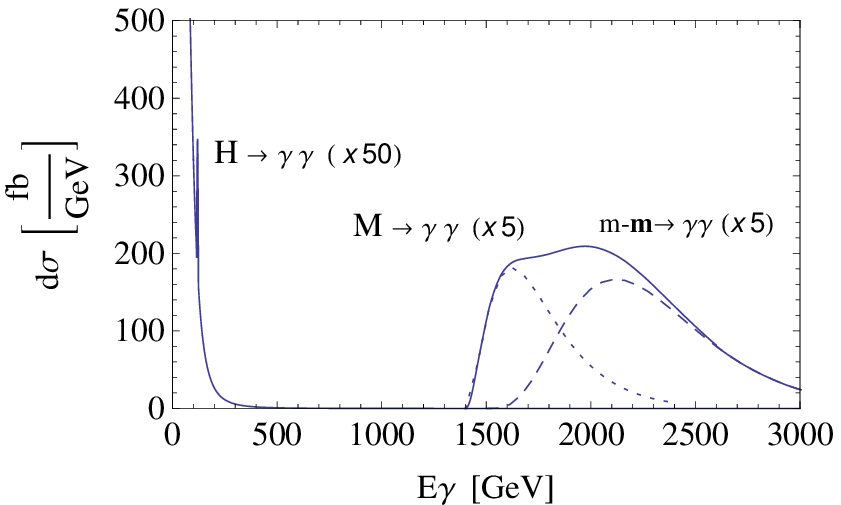}
\caption{We show the differential two photon crossection with the Higgs signal (scaled by a factor of 50 over the background) where the monopolium and monopole-antimonopole contributions have been incorporated. We also show the monopolium signal (dotted)  and monopole- antimonopole as in ref. \cite{Epele:2011cn} (dashed) (scaled by a factor of 5). The mass of the monopole in both cases is $750$ GeV and that of the monopolium $1400$ GeV. Note that for $1 fb^{-1}$ the y-axis measures counts in $1$ GeV bins. The background has been obtained from ref. \cite{atlas1}. }
\label{mmMhiggs}
\end{center}
\end{figure}

Two are the main physical dependences  of the cross section: the monopole mass, $m$,  and the monopolium mass , $M$. In Fig. \ref{MggEbind} we fix the monopole mass to $750$ GeV and vary the binding energy. We see that the cross section increases dramatically with binding energy. Thus the effect of the binding is twofold: it increases the cross section and it lowers the threshold from $2m$ the monopole-antimonopole production threshold. To observe the other dependence, in Fig. \ref{Mmonopolemasses} we fix the binding energy to $100$ GeV and vary the monopole mass. The effect goes inversely proportional to the monopole mass, i.e., the lower the monopole mass the larger the  cross sections.

We can summarize our findings by stating  that low monopole masses and large bindings favor the detection of monopolium. Monopolium has the advantage over the monopole-antimonopole process studied in ref \cite{Epele:2011cn} of lowering the threshold, narrowing the bump and  increasing the cross section with binding energy. In Fig. \ref{mmMhiggs} we show three interesting effects, namely the two photon decays of Higgs,  monopolium and monopole-antimonopole annihilation. The Higgs signal has been increased over the background by a factor of 50. The parameters for the monopolium cross section shown are $m=750$ GeV and $ M= 1400$ GeV, which have been chosen so that its signal is of the same size of that of the monopole-antimonopole annihilation as calculated in ref.  \cite{Epele:2011cn} with a monopole mass of also $750$ GeV.  We note  two  of the features mentioned before, the lower threshold and the narrower bump structure. If we would increase the binding by a few tens of GeV the height of the bump would increase considerably with respect to the monopole-antimonopole cross section (recall Fig.\ref{MggEbind}). Note that the $\gamma \gamma$ conventional background at the monopole scenario is extremely small, as measured recently by ATLAS in di-photon studies
\cite{diphoton} and while searching for $H\rightarrow\gamma\gamma$~\cite{hgg}.

If  a broad bump would appear experimentally one should confirm that it could arise from monopole dynamics. Thereafter  the way to distinguish between the decay of monopolium and the annihilation of monopole-antimonopole would be through the angular
dependence. In the present scenario, a spin zero monopolium, the angular decay properties would be similar to that of para-positronium \cite{Adkins:2001zz}, while that of monopole-antimonopole would be analogous to light-by-light scattering in QED \cite{Karplus:1950zz,Csonka:1974ey}. Moreover, the two pehenomena could occur simultaneausly, as happens in the case of electrons and positrons, where we have in light-by-light scattering electron-positron annihilation and positronium decay. If the latter were the case two bumps, if the overlap is not large, or a very broad flat bump, if the overlap is considerable, could be seen. The existence of one or two bumps depends very strongly on the binding dynamics and the monopole mass. 
Note that there is no possible confusion with the Higgs, since its width is narrow compared to its mass. Moreover, the scattering cross section
for a heavy Higgs in the two photon channel is extremely small compared to that in the other channels, and therefore, its characteristics would be known by the time two photon bump
would be seen.

\section{Conclusions}

The Dirac quantization condition is a beautiful consequence of the existence of monopoles and therefore they represent an extremely appealing physical scenario. 
There is as of yet no experimental proof of their existence. This has led to approximate mass bounds which suggest a mass scale for the monopole 
above 500 GeV. LHC opens up this energy regime for research and therefore monopoles become again a subject of experimental search.

Even if monopoles exist it might be possible that due to the very strong magnetic coupling they do not appear as free states but bound forming 
monopolium, a neutral state,  very difficult to detect directly. Here we have  analyzed the coupling of monopolium to photons and its contribution 
to light-by-light like scattering, i.e. monopolium's dynamical decay. We have found that for reasonable values of the monopole mass and relatively 
small, compared with their mass, binding energies, spin zero monopolium desintegrates into two gammas with cross sections which are reachable in this first run of LHC. 

Our investigations go beyond this wishful scenario. We have seen that the cross section depends both on the mass on the monopole and on that 
of monopolium, increasing inversely proportional to the monopole mass and being directly proportional to the binding energy. This means that 
similar cross section can be achieved with very heavy monopoles if the binding energy is large. This is however a qualitative statement since our calculation has been carried out in a non relativistic framework for which large bindings are not well under control.

To conclude, the eventual appearance of broad bumps in the two photon cross section might be associated with the existence of monopoles, either free, as discussed in our previous work~\cite{Epele:2011cn}, and/or bound in monopolium as presented here. The implementation of our findings in detector analysis would provide actual observations.

\section*{Acknowledgement}
We thank the authors of JaxoDraw  for making drawing diagrams an
easy task \cite{Binosi:2003yf}.  LNE, HF
and CAGC were partially supported by CONICET and ANPCyT Argentina. VAM acknowledges support by the Spanish Ministry of Science and Innovation (MICINN) under the project FPA2009-13234-C04-01, by the Spanish Agency of International Cooperation for Development under the PCI project A/030322/10 and by the grant UV-INV-EPDI11-42955 of the University of Valencia. VV has been supported  by HadronPhysics2,  by  MICINN (Spain) grants FPA2008-5004-E,  FPA2010-21750-C02-01,   AIC10-D-000598 and by GVPrometeo2009/129.

\end{document}